\documentclass[twocolumn,aps,showpacs,amsmath,preprintnumbers,floatfix,a4paper,prl]{revtex4}

\usepackage{graphicx}
\usepackage{dcolumn}
\usepackage{bm}
\usepackage{color}
\usepackage{amssymb}
\usepackage{natbib}
\usepackage{times}

\begin{document}

\title{Radially Polarized, Half-Cycle, Attosecond Pulses from Laser Wakefields
through Coherent Synchrotron Radiation}

\author{F. Y. Li$^{1}$}
\author{Z. M. Sheng$^{1,2}$}
       \thanks{zmsheng@sjtu.edu.cn or zhengming.sheng@strath.ac.uk}
\author{M. Chen$^{1}$}
\author{L. L. Yu$^{1}$}
\author{J. Meyer-ter-Vehn$^{3}$}
\author{W. B. Mori$^{4}$}
\author{J. Zhang$^{1}$}

\affiliation{$^1$ Key Laboratory for Laser Plasmas (MoE),
Department of Physics and Astronomy, Shanghai Jiao Tong University, Shanghai 200240, China\\
$^2$ SUPA, Department of Physics, University of Strathclyde, Glasgow G4 0NG, UK\\
$^3$ Max-Planck-Institut f\"{u}r Quantenoptik, D-85748 Garching, Germany\\
$^4$ University of California, Los Angeles, California 90095-1547, USA}

\date{\today}

\begin{abstract}
Attosecond bursts of coherent synchrotron-like radiation
are found when driving ultrathin relativistic electron disks
in a quasi-one-dimensional regime of wakefield acceleration,
in which the laser waist is larger than the wake wavelength.
The disks of overcritical density shrink radially due to
the focusing wake fields, thus providing the transverse currents
for the emission of an intense, radially polarized, half-cycle
pulse of about 100 attoseconds in duration. The electromagnetic
pulse first focuses to a peak intensity 10 times larger
($7\times10^{20}\rm W/cm^2$) than the driving pulse and
then emerges as a conical beam. Saturation of the emission
amplitudes is derived analytically and in agreement with
particle-in-cell simulation. By making use of gas targets
instead of solids to form the ultrathin disks, the new
scheme allows for high repetition rate required for applications.
\end{abstract}

\maketitle

Great efforts have been devoted to obtain extreme-ultraviolet
(XUV) and soft X-ray pulses having attosecond duration.
They open the door to attosecond spectroscopy of bound
electrons~\cite{Krausz2009}. An established method to produce
such pulses relies on atomic harmonics excited by ultrafast
lasers at an intensity below $10^{16}~\rm W/cm^2$~\cite{gasHHG}.
Harmonics from solid surfaces~\cite{Teubner2009} make use
of relativistic optics ($>10^{18}~\rm W/cm^2$)~\cite{Mourou}
and allow for much brighter sources. Single attosecond spikes
can be isolated from trains of pulses by means of high-pass
filters~\cite{HHG}. By temporally rotating either the surface
geometry~\cite{Naumova2004} or the driver wavefront
~\cite{Wheeler2012}, individual attosecond pulses
can be obtained even without filtering.

In this Letter, we follow a different path, producing
intense isolated attosecond pulses from underdense
plasmas rather than solids. With gas targets
the condition of ultrahigh laser
contrast is greatly relaxed, and a bright
source allowing for high repetition rate may become possible.
The new path is based on wakefield acceleration
driven by an intense short laser pulse.
As it is well known, these relativistic plasma waves
exhibit strong accelerating and focusing fields for
electrons. They promise acceleration to high energies
over a short distance~\cite{LWFA}.
Often the laser beam is tightly focused to reach
highest possible intensity $I_0$ with waist
$W_0\leq\sqrt{a_0}\lambda_p/\pi$, where $\lambda_p$
is the plasma wavelength and
$a_0=8.5\times10^{-10}\lambda_0[\rm\mu m]\sqrt{I_0[W/cm^2]}$
the normalized laser amplitude at wavelength $\lambda_0$.
Plasma electrons are then pushed sideways by the
light pressure, creating bubble-like wakes.
Some electrons circling around the bubble are transversely
injected at the rear vertex and then accelerated
forming a narrow bunch~\cite{Kostyukov2009PRL}.
This is known as the bubble regime of wakefield
acceleration~\cite{Bubble}. An important feature is that
the accelerating bunch oscillates in the wake forced by
the focusing fields and emits bright betatron
radiations~\cite{Betatron}.

Here we make use of a different regime of wakefield acceleration,
occurring for laser waists larger than wake wavelength.
Wake electrons now perform mainly one-dimensional (1D)
longitudinal motion due to weak radial expelling.
The density wave crests show disk-like profiles as visualized in
experiments~\cite{Matlis2006}. Most notably, they compress
into dense sheets when driven close to wavebreaking.
Actually such ultrathin wave crests have
been used as relativistic mirrors to backscatter
femtosecond probe pulses~\cite{FlyingMirror}.
In this way, attosecond XUV/X-ray pulses are obtained
owing to simultaneous frequency upshift and pulse
shortening by Dopper factors of $\sim4\gamma_p^2$,
where $\gamma_p=1/\sqrt{1-v_p^2/c^2}$ with $v_p$
the wake's phase velocity and $c$ the light speed
in vacuum. In this work, however,
we follow another line that does not require
a second pulse, but leads to self-emission of a
bright attosecond pulse from the dense electron sheet itself.
For this to happen, the electron sheet has to be
injected and accelerated in the wakefield.
Different from the ultrathin density crest belonging
to the quasi-1D waveform~\cite{FlyingMirror},
the injected sheet contracts in transverse direction
due to the focusing wake fields, while boosted
at the same time in energy by wakefield acceleration.
The central new result of this Letter is that
a strong unipolar attosecond pulse is produced
by the transverse currents formed during the
contraction motion. This coherent attosecond source
is also in contrast to the incoherent
femtosecond betatron X-rays normally
obtained so far~\cite{Betatron}.

Different ways may be used to induce injection of
such electron sheets~\cite{Nanosheet}.
A simple method has been described
using an up-ramp density profile followed
by a plateau~\cite{Li2013prl}.
Wavebreaking then occurs at the transition point
and leads to a sudden longitudinal injection into
the quasi-1D wake, different from the continuous
transverse injection in the bubble regime~\cite{Kostyukov2009PRL,Bubble}.
The key point is that, along the ramp, the wave crest travels
at superluminal speed, preventing premature injection.
At the transition to the plateau, the density spike
is injected as a whole due to fast switching of $v_p$.
It forms an ultrathin (few nm) overcritical dense
electron disk that accelerates in the wakefield.

\begin{figure}[t]
\centering
\includegraphics[width=0.45\textwidth]{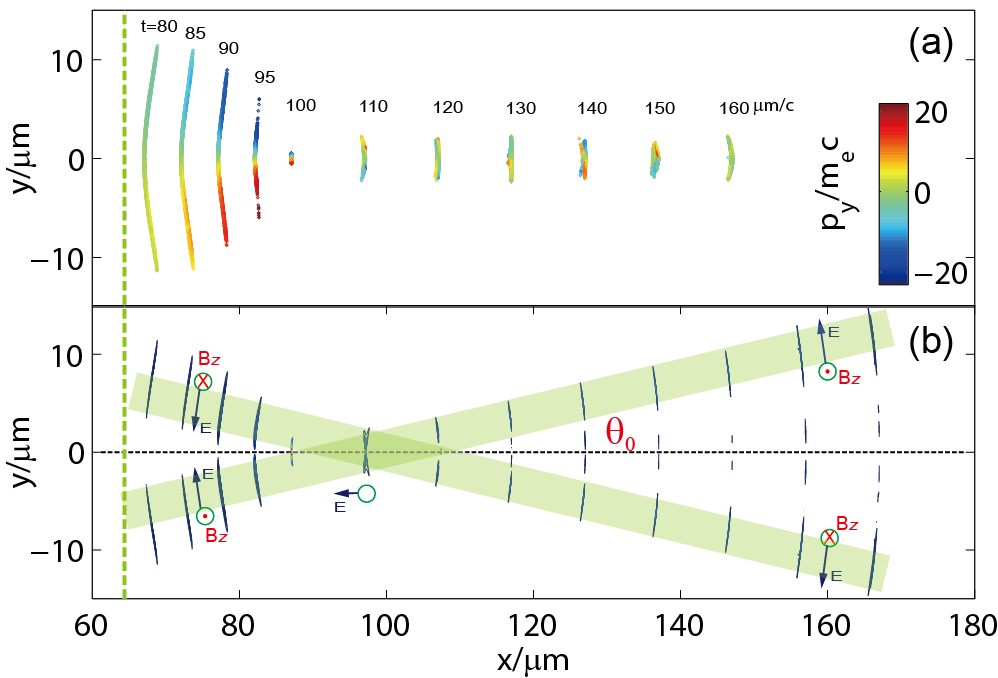}
\caption{(Color online).
(a) Snapshots of trapped electron sheet at different times
(given as label) and colored according to transverse momentum
$p_y$ and (b) the radiated attosecond emission given as
contour plot of $B_z^2$. The dashed line on the left
side marks the point of wavebreaking and injection into
the wake. The full density pattern of the wake is
given in Fig. 2(b) for $t=160~\rm\mu m/c$.
These results are obtained from 2D PIC simulation
in $x$-$y$ space. In full 3D geometry the emission
appears as an annular beam propagating as a radially
polarized, half-cycle, attosecond (RHA) pulse along a
cone with opening angle $\theta_0\sim10^\circ$,
indicated as shaded stripes.}
\end{figure}

The evolution of an electron sheet in the wakefield and
the associated radiation flash are first illustrated
in Figs. 1(a) and 1(b). It is seen that the sheet
contracts transversely to a smallest diameter
in tens of laser periods after injection, while accelerating
longitudinally up to a highly relativistic $\gamma$-value
(compare Fig. 3). It is in this short time interval that
the attosecond pulse is emitted. As found in Fig. 1(b),
the pulse is first attached to the electron sheet,
but then propagates along a cone, while the electrons
move closer to the axis in a channel of almost constant diameter.

These results are obtained from two-dimensional particle-in-cell
(2D PIC) simulations~\cite{Fonseca2002}.
Here in order to control trapping of the electron sheet,
the self-injection method with particular density tailoring
was employed~\cite{Li2013prl}. A linearly
polarized (along $z$ direction), $18~\rm fs$ (full width
at half maximum) laser pulse of peak intensity
$I_0=7.7\times10^{19}~\rm W/cm^2$
irradiates an underdense plasma slab with a
$65\rm\mu m$-long ramping front edge.
The plateau density is $n_0=0.04n_c$ with
$n_c=1.74\times10^{21}~\rm cm^{-3}$ the critical
density for $\lambda_0=0.8~\rm\mu m$.
A laser waist of $W_0=17~\rm\mu m$ is enough for
the formation of wide sheets in the quasi-1D regime.
In the simulation, $158\times25$ grids cells
per $\rm\mu m^2$ were used and sufficient to
resolve the attosecond spikes; even higher resolutions gave
almost the same results. Furthermore, an initial electron
temperature of $T_e=20~\rm eV$ was chosen to
mimic field ionization by prepulse.

We now discuss the radiation process in general.
The injected electrons are diverted by the focusing wake
fields toward the central axis according to
$d\beta_{\perp}/dt\simeq \varepsilon_{\perp}/\gamma$.
Here $\beta_{\perp}$ is the normalized transverse velocity;
$t$ and the focusing field $\varepsilon_{\perp}$ have,
respectively, been normalized by
$\omega_0^{-1}$ and $E_0=m_e\omega_0c/e$ with electron
mass $m_e$, elementary charge $e$ and laser frequency
$\omega_0=2\pi c/\lambda_0$.
Near the central axis, the focusing field,
$\varepsilon_{\perp}\propto n_0r$,
is almost linear in radius $r$ and the ambient density
$n_0$. Trapped electrons then perform synchrotron-like
motion with curvature radius
$\rho\simeq\lambda_0\gamma/(2\pi \varepsilon_{\perp})$
and emit broadband radiation with cutoff frequency
$\omega_c\simeq3\gamma^3c\rho^{-1}$~\cite{Shiozawa}.
However, there are a couple of new features when compared
with bubble betatron radiation~\cite{Betatron}.
First, the high density of the sheet,
typically larger than $10^{21}~\rm cm^{-3}$~\cite{Li2013prl},
enables synchrotron emission in a coherent manner.
Coherence occurs provided that a sufficiently large
number of electrons resides in a volume with scale length
equal to the radiation wavelength in the rest frame of
electrons. For the present case, it requires the
sheet density to satisfy
$n_s\gg10^{13}\varepsilon_{\perp}^3\gamma^4~\rm cm^{-3}$,
where radiation at the cutoff wavelength
$\lambda_c=2\pi c/\omega_c$ is assumed.
Considering $\varepsilon_{\perp}\sim 0.1$,
this criteria is readily met even for $\gamma>100$.
Second, due to axial symmetry and inward acceleration
of the disk electrons, a radially polarized half-cycle
electromagnetic pulse is emitted.
Third, with the increase of $\gamma$ by wakefield acceleration,
the radiated power grows $\propto\gamma^2$~\cite{Shiozawa}.
That means the dominant emission profile roughly
overlaps the sheet during amplification,
having an attosecond duration.
It is worth noting that
similar radially polarized pulses have actually been
detected from thin photoconductors, where annular
microelectrodes induced the radial current~\cite{Kan2013}.
Due to the static wafer plane, however, only pulses of
picosecond duration were observed, determined by
the lifetime of the radiating current.

These features are well illustrated in the sample simulation.
Figure 2(a) records how the peak field
$B_z^{max}/B_0$ of the radially polarized, half-cycle,
attosecond (RHA) pulse evolves; $B_0=m_e\omega_0/e$ is the
normalizing field. According to Fig. 1(b),
the pulse emerges in three stages:
pulse generation during initial sheet contraction,
beam crossing on axis, and finally annular beam propagating
along a cone. At the phase of crossing, the emission focuses
to a maximum of $B_z^{max}/B_0\geq18$, as found in the
shaded region of Fig. 2(a). This corresponds to a peak
intensity greater than $7\times10^{20}~\rm W/cm^2$.
After complete separation ($t\sim130~\rm\mu m/c$),
the actual peak field shows up as $B_z^{max}/B_0=4.5$,
corresponding to $4.3\times10^{19}~\rm W/cm^2$.
Importantly, Fig. 2(b) shows how the RHA pulse
propagates at later time. The initial quasi-1D wake
is evolving into a complex broken-wave pattern due to laser
evolution~\cite{Xujc2007}. The half-cycle pulse as
short as $120~\rm as$ (see Fig. 2(c)) is propagating
inside the wake but deviated from the axis.
It contains a broadband XUV spectrum extending
up to $\sim 40~\rm eV$ in photon energy.
In the present case, the annular
beam carries a total energy of $12~\rm mJ$,
which is $\sim2\times10^{-3}$ of the incident laser energy.
The radial polarization of the attosecond pulse
is confirmed in Fig. 2(c), where switching
polarization of the driving pulse from
$y$ to $z$ direction has almost no effect on
the emission profile. We have checked that the
annular beam is uniform azimuthally when
obtained in 3D simulation.

\begin{figure}[t]
\centering
\includegraphics[width=0.45\textwidth]{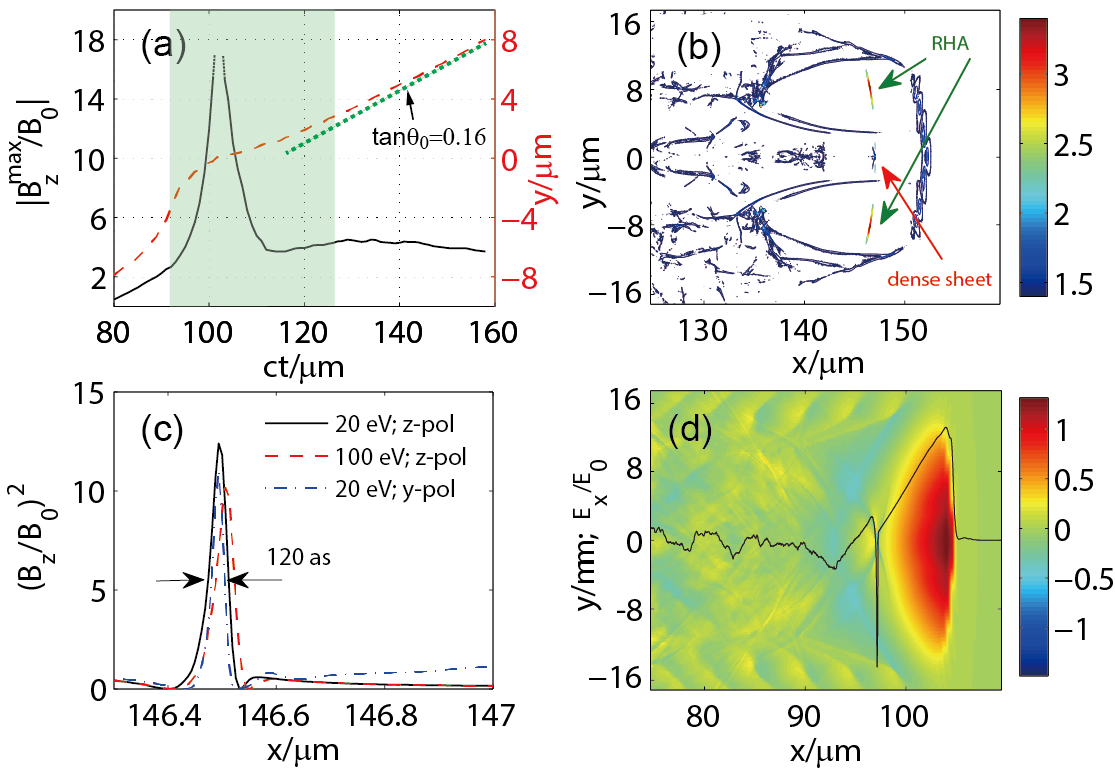}
\caption{(Color online). (a) Evolution of peak field
$|B_z^{max}|/B_0$ (solid) and associated $y$ coordinate
(dashed) of the RHA pulse versus time. The shaded region
refers to the crossing stage as explained in the text.
The green dotted line indicates the slope of propagation direction.
(b) Contour plots of the electron density $n/n_c$ plus $|B_z/B_0|$
(share the same color bar) at $t=160~\rm\mu m/c$.
At this late time, the wake has developed a complex pattern,
in particular to the sides bordering unperturbed plasma
and downstream. The only features relevant to this Letter
are the tiny dense sheet of relativistic electrons
(marked by red arrow) and the RHA pulse
(marked by green arrows). (c) Line-outs of
$(B_z/B_0)^2$ along $y=8.32~\rm\mu m$ at $t=160~\rm\mu m/c$ for
three cases corresponding to different initial plasma temperatures
(20 and 100$~\rm eV$) and $y$ or $z$ polarization of driving laser
pulse. (d) Snapshot of the longitudinal electric field $E_x$
at $t=110\rm\mu m/c$; black curve ($\times$10) shows line-out along
central axis.}
\end{figure}

The basic dynamics of RHA generation can be derived
from a simplified model. It assumes that a flat,
monoenergetic electron sheet of delta-like density profile
$n_s=\delta(x/\sigma_s)$ with finite areal density
$\sigma_s=\int n_sdx$ and no initial transverse momentum is
injected into a wakefield of uniform accelerating
and focusing fields. Coherent radiation from the
electron sheet is described by the 1D wave equation
$(\partial^2/\partial x^2-c^{-2}\partial^2/\partial t^2)E_{y,r}
=\epsilon_0^{-1}c^{-2}\partial J_{y,r}/\partial t$,
where $J_{y,r}=-ec\beta_{y,r}n_s$ is the
radiating current with
$\beta_{y,r}=\int_{\tau_0}^{\tau_{1}} d\beta_y\ll1$
the velocity integrated over a short transverse
acceleration~\cite{Wuhc2012}. The radiated field
$E_{y,r}$ is first calculated in the rest frame of
the sheet with Lorentz factor $\sim\gamma$ and
then transformed to the laboratory frame;
the result is given by
\begin{equation}
    \label{eq1}
E_{y,r}\simeq(\frac{\sigma_s}{\epsilon_0})\beta_{y,r}^R\gamma,
\end{equation}
where $\beta_{y,r}^R$ denotes the velocity $\beta_{y,r}$
in the rest frame. Equation~(\ref{eq1}) indicates that
the radiated power grows $\propto|E_{y,r}|^2\propto\gamma^2$,
while $\gamma$ follows from
$d\gamma/dt=-\vec{\beta}_x\cdot\vec{E}_x-\vec{\beta}_y\cdot\vec{E}_y$.
Taking into account the radiated field $E_{y,r}$,
the total fields can be expressed as
$\vec{E}_x=\vec{E}_{x,w}$ and
$\vec{E}_y=\vec{E}_{y,w}+\vec{E}_{y,r}$,
where $\vec{E}_{x,w}$ and $\vec{E}_{y,w}$ are the
longitudinal and transverse wake fields, respectively.
Inserting the expression of Eq.~(\ref{eq1}),
the energy equation reads
\begin{equation}
    \label{eq2}
\frac{d\gamma}{dt}+\frac{\sigma_s}{\epsilon_0}\beta_{y,r}^R\beta_y\gamma
-\beta_yE_{y,w}-\beta_xE_{x,w}=0.
\end{equation}
Assuming $\gamma(t=0)=1$ initially, wakefield acceleration
first boosts the $\gamma$-value rapidly, which then
saturates due to radiation damping at a rate
$(\sigma_s/\epsilon_0)\beta_{y,r}^R\beta_y$.
Saturation occurs for
$\vec{\beta}_x\cdot\vec{E}_x+\vec{\beta}_y\cdot\vec{E}_y=0$,
leading to the saturated field
\begin{equation}
    \label{eq3}
E_{y,r}^{sat}\simeq(E_{x,w}/\tan\theta+E_{y,w}),
\end{equation}
where $\tan\theta=\beta_y/\beta_x$ is evaluated at
the saturation instant.

It is clear that wakefield acceleration has played an
essential role in the attosecond pulse generation.
This marks a most prominent feature of the present scheme.
Equation~(\ref{eq3}) suggests that $E_{y,r}^{sat}$
can be large provided that $\tan\theta\simeq\beta_y$
is small initially and the saturated level can be
maintained for a long period. There are a couple of
other effects not yet included. For the sample case,
the injected sheet is not perfectly flat but slightly
curved as seen in Fig. 1(a). Due to the
sudden longitudinal injection, a curved wake
wavefront~\cite{FlyingMirror,Matlis2006} directly maps
into the injected sheet profile and the sheet electrons
also inherit finite transverse velocities.
These will cause their transverse contraction even without
the focusing fields, and the radiation process
is ultimately limited by deformation of the sheet profile.

\begin{figure}[t]
\centering
\includegraphics[width=0.45\textwidth]{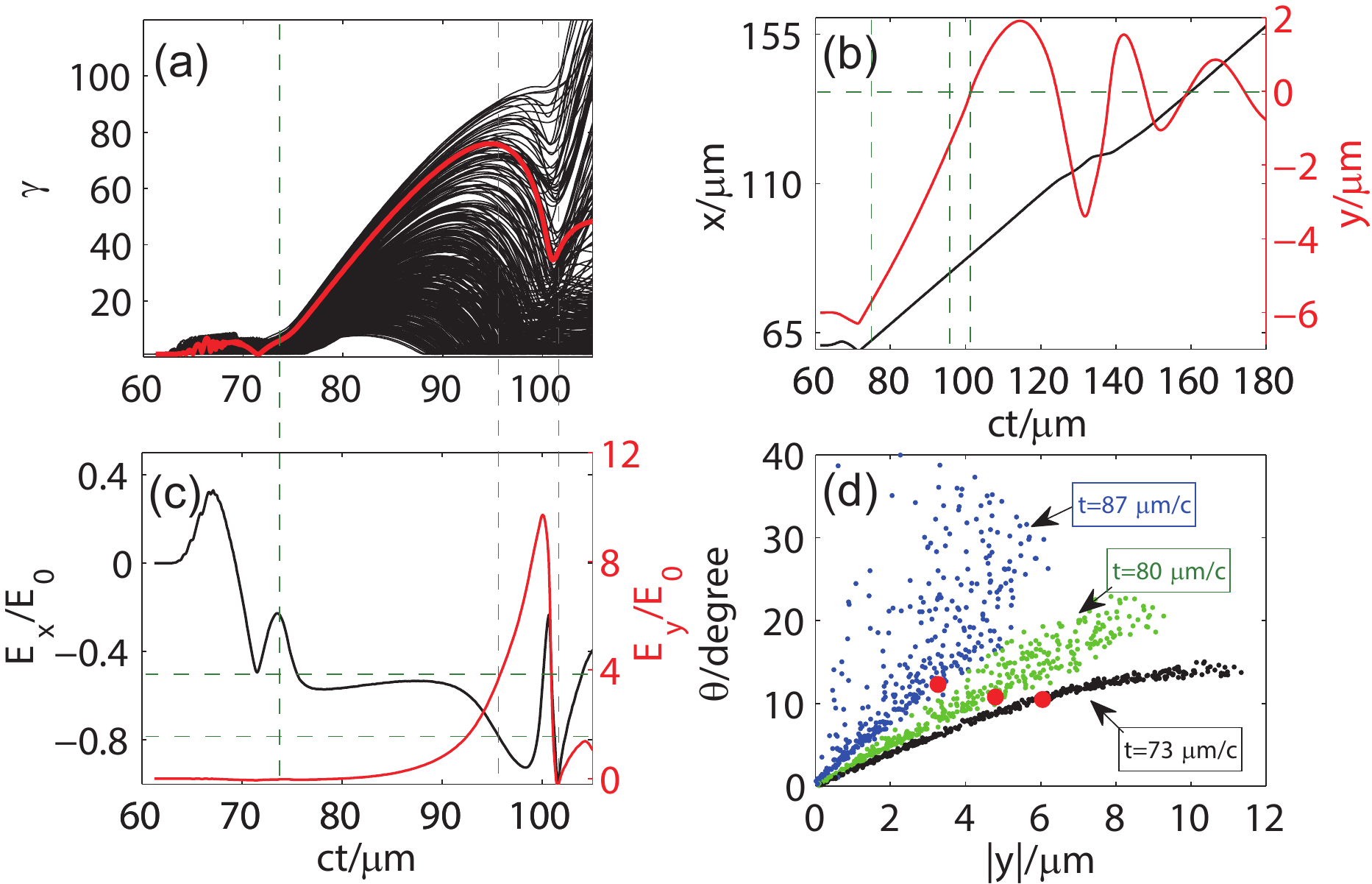}
\caption{(Color online). Particle tracking of 400 uniformly distributed
electrons picked from the lower half of the sheet at $t=80\rm\mu m/c$.
(a) Their $\gamma$ factors versus time. The red curve highlights a typical
electron which was initially located near the center of the lower half
and belongs to the most energetic ones after acceleration.
Time histories are presented for this electron in terms of
(b) its coordinates and (c) the electric fields $E_x$ and
$E_y$ felt by it. The vertical dashed lines (from left to
right) marks, respectively, the time of injection,
$\gamma$ saturation and reaching $y=0$.
(d) Distributions of the propagation angle
$\theta\simeq(180^\circ/\pi)\arctan\beta_y$
for these electrons versus $|y|$ at the times indicated;
$t=73~\rm\mu m/c$ corresponds to just before injection.
The red dots refer to the typical electron defined in (a).}
\end{figure}

Let us explore these effects by particle tracking.
Figure 3 presents the results for a group of electrons
uniformly distributed in the lower half of the electron sheet.
According to Figs. 3(b) and 3(c), the sheet is injected at
$t\simeq74~\rm\mu m/c$ and focuses to the central axis at
$t\simeq102~\rm\mu m/c$. During contraction
these electrons develop a broad $\gamma$ spread
due to high thermal temperature; see Fig. 3(a).
However, the overcritical density and ultrathin
feature of the sheet ensure the coherence of the
attosecond pulse. It is hardly affected by
initial plasma temperatures; results given for
$T_e=10~\rm eV$ and $100~\rm eV$
in Fig. 2(c) show little difference.
Moreover, as found in Fig. 3(d), the transverse velocities
$\beta_y$ of the sheet, inherited from wave crest,
scale almost linearly with $|y|$
(details see supplemental material)
and increase with time due to transverse acceleration.
This suggests that the sheet can be subdivided into
a sequence of ring-shaped segments defined by
their particular $\beta_y$. Applying the simplified model
to each segment with coordinate frame rotated
by angle $\theta\simeq\arctan\beta_y$ from the normal
$x$-$y$ axis, the resulting pointing direction of the
radiated electric field will deviate from the $y$ axis;
see the schematic drawing in Fig. 1(b).
A clear evidence for this is shown in Fig. 2(d),
where a sharp negative $E_x$ field shows up
at the on-axis overlap joint during beam crossing.
Emissions from segments of different $\beta_y$ then
converge due to propagation. This accounts for the
focusing effect observed in Fig. 2(a).
The finally observed RHA pulse (Fig. 2(b))
builds up from all segment contributions weighted
according to Eq.~(\ref{eq1}). As it turns out,
largest contributions stem from ring segments
with medium radius. For the present case,
the conical angle $\theta_0$ of
the radiation peak is measured to be
$\tan\theta_0\simeq0.16$ in Fig. 2(a).
To evaluate this peak field, a typical electron
is highlighted in Fig. 3; it is initially located near
the center of the lower half and belongs to the most
energetic ones after acceleration.
According to Eq.~(\ref{eq3}), the peak field amounts to
$E_{y,r}^{sat}\simeq E_x/\tan\theta_0=4.7$,
which is in fair agreement with the observation of
$B_z^{max}/B_0=4.5$ after $t\sim130~\rm\mu m/c$.
Here $E_x\simeq0.75$ is the longitudinal electric field felt
by the electron at the saturation instant and
is extracted from Figs. 3(a) and 3(c).

\begin{figure}[t]
\centering
\includegraphics[width=0.43\textwidth]{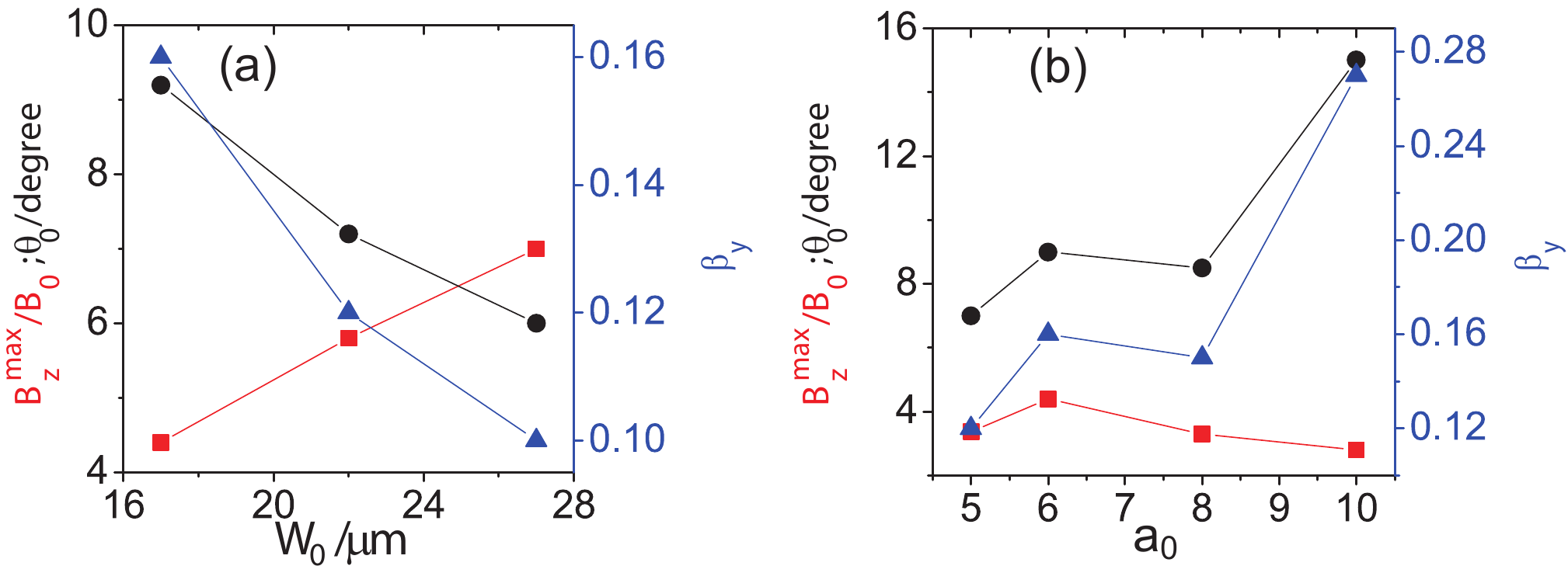}
\caption{(Color online). Peak radiation field $B_z^{max}$
(solid square), $\theta_0$ defined in Fig. 1 (solid circle),
and $\beta_y$ of the sublayer emitting the peak field
(solid triangle) versus (a) laser waist $W_0$,
while keeping $a_0=6$ and $n_0=0.04n_c$ fixed,
and versus (b) driver amplitude $a_0$,
while keeping $W_0=17\rm\mu m$ and $n_0=0.04n_c$ fixed.}
\end{figure}

The present RHA generation is robust against changing
laser and plasma parameters, provided that they are
sufficient to induce the sheet injection.
Larger focal spots of the driving pulse generally lead to
smaller initial transverse velocities of the injected sheet,
generating higher radiation peaks following Eq.~(\ref{eq3}).
This is shown in Fig. 4(a) where $B_z^{max}$ and
$\theta_0$ scale almost linearly with $W_0$.
While scanning the laser amplitude $a_0$ for fixed
plasma density $n_0=0.04n_c$, an optimal value of
$a_0\sim6$ exists as seen in Fig. 4(b).

Petawatt lasers, presently coming up, will provide sufficiently
high power for the high intensities and the relatively
broad focal spots required here. The sample case shown
above used a single $350~\rm TW$ driving pulse delivering
$6.5~\rm J$ in $\sim20~\rm fs$. These parameters are well
within the ELI-facility capabilities~\cite{ELI} and are
also becoming available commercially. The conical angle
of the RHA beam is typically less than $10^\circ$,
thereby allowing refocusing even at half-meter distance
by a concentric mirroring tube with diameter less than
$16~\rm cm$; such techniques have actually been implemented
in the measurement of wakefield-based electro-optic
shocks~\cite{Helle2010PRL}. The RHA pulse should also
be easily distinguished from other sources due to its
radial polarization~\cite{Schnell2013} and hollow pattern.

In conclusion, we have identified novel attosecond bursts
of coherent synchrotron radiation from laser wakefield
acceleration. The attosecond feature does not derive
from an ultrathin solid foil~\cite{Wuhc2012} or surface
layer~\cite{Nanobunch}, but arises from intrinsic
features of nonlinear plasma waves, namely steepening
and breaking. It makes use of quasi-1D wavebreaking,
allowing to trap ultrathin electron sheets in the wakefield.
The electron sheet contracts, while boosted in energy,
and is found to emit a relativistically intense
attosecond pulse. Pulse energy exceeding $10~\rm mJ$
can be obtained with the laser-to-RHA conversion efficiencies
beyond $10^{-3}$. Besides being a bright source
for attosecond applications, this burst will also
provide excellent diagnostics for the
wavebreaking dynamics~\cite{Thomas2007prl}
in the quasi-1D regime and corresponding
electron sheet generation~\cite{Li2013prl}.

ZMS would like to thank the OSIRIS Consortium
at UCLA and IST (Lisbon, Portugal) for providing access
to OSIRIS 2.0 framework. This work is supported in part by the
National Basic Research Program of China (Grant No. 2013CBA01504),
the National Science Foundation of China (Grant No. 11121504, 11205101,
11374209 and 11374210), the MOST international collaboration
project (Grant No. 2014DFG02330), and the EPSRC, UK.
Simulations were performed on Magic Cube at Shanghai Supercomputer Center
and $\Pi$ at Shanghai Jiao Tong University.

\end{document}